\documentclass{JHEP3}

\usepackage{graphicx}
\usepackage{amssymb}
\usepackage{amsbsy}
\usepackage{makeidx}

\makeindex

\newcommand{\half}{\frac{1}{2}}

\renewcommand{\d}{\delta}

\newcommand{\dd}[2]{\frac{\partial #1}{\partial #2}}
\newcommand{\nn}{\nonumber}

\newcommand{\s}{\phantom{a}}

\newcommand{\ep}{\epsilon}

\newcommand{\be}{\begin{eqnarray}}
\newcommand{\en}{\end{eqnarray}}

\renewcommand{\vss}{\vspace{7mm}}

\renewcommand{\d}{\partial}

\title{Localizing Gravity on the Triple Intersection of 7-branes in 10D}
\author{A. Liam Fitzpatrick and Lisa Randall\\ Jefferson Physical Laboratory, Harvard University, Cambridge, MA, 02138 \\ E-mail: \email{fitzpatrick@physics.harvard.edu},\email{randall@physics.harvard.edu}}
\abstract{
It was recently proposed  that our universe could naturally come to be dominated
 by 3-branes and 7-branes if the universe is ten-dimensional.  In this paper, we explicitly  demonstrate that gravity
 can be localized on the intersection of three 7-branes in AdS$_{10}$ to give four-dimensional gravity.
We derive the exact relations among the
 tensions of the branes, and show that 
they apply independently of the precise  distribution of energy within the  necessarily thickened  branes.
 We demonstrate this with  several technical  sections  showing
a simple formula for  the curvature tensor  of a  diagonal metric with isometries  as well as
 for the curvature at a gravitational singularity.
We also demonstrate a subtlety in applying Stoke's Theorem to this set-up.
}

\preprint{.}

\begin{document}

\vspace{-2mm}

\section{Introduction}

Even though a theory with extra dimensions must allow gravitons to
propagate in all directions,  
four-dimensional gravity can apply even with infinitely large extra dimensions if
 they are sufficiently warped \cite{rsii}. There are many ways of understanding
this result but from the
four-dimensional perspective, the higher-dimensional graviton will
appear as a tower or spectrum of four-dimensional graviton fields
with different masses, similar to the usual Kaluza-Klein case.
The spectrum is gapless and continuous and contains a normalizable zero mode (or almost zero mode) that
dominates the gravitational potential.  While this mechanism is
simplest for a single codimension-one brane, it has been
generalized to higher codimension \cite{gs,inflargedims,gstwo}.
Such setups are appealing
 since string theory motivates a ten-dimensional spacetime, which would make the visible
 universe a codimension-six brane.

It was recently pointed out \cite{relaxing} that a generic ten-dimensional FRW cosmology would be dominated
by 3-branes and 7-branes.
 While a 3-brane in this universe would not generally exhibit 4D gravity, the intersection of three 7-branes might.
 Each 7-brane is codimension-2 and can localize gravity to
  itself \cite{gs}.  Intuitively, then, gravity might be localized to the intersection, 
as Ref. \cite{inflargedims} showed for the codimension-1 case. Moreover, we
   generically expect three 7-branes to intersect over a four-dimensional spacetime surface in ten dimensions.
     
  In this  paper , we focus on the triple 7-brane intersection and show that it can
localize four-dimensional gravity.  
We first construct our solution explicitly and then, using the high degree of symmetry of our construction, 
demonstrate how to extract the tension relations about the thickened branes  in terms of the known external metric and a few parameters of the interior metric of the brane.  
In particular, we find the necessary tuning relations and show 
that for a flat four-dimensional universe we do not require an extra tensionful brane at the intersection. 
 We calculate the graviton potential and show that gravity is localized on the intersection.  In an appendix, we present 
the explicit construction for the same setup with AdS$_4$ or dS$_4$ on the intersection 
and calculate the leading cosmological constant (c.c.)-dependent term.   

It might seem surprising that we can find exact tension relations for codimension-2 branes and their intersections,
given that  the codimension-2 branes should be treated as thick branes, and you would expect the tension
relation to depend on the precise form of the metric on the interior. However, we will demonstrate that one can apply 
Stoke's theorem to relate the AdS curvature to the tension on the boundary. Our calculation in fact generalizes the
 surprising fact already seen in 
\cite{gs,gregory, navarro} that the tension relations depended only on boundary conditions and not on the detailed form of the interior
metric of a codimension-2 brane.  There is a subtlety in that we also
need to take into account an interior contribution which amounts to an internal surface that depends on only a few 
boundary condition parameters.  To apply  Stoke's Theorem, we need to account for the curvature at a singularity and we show how to do this in the text.

\section{Review of the Gherghetta-Shaposhnikov Construction}
The authors (GS) of \cite{gs} demonstrated that  gravity can be localized on a codimension-2 brane. They considered a codimension-2  Minkowski 4-brane embedded in AdS$_6$,
whereas our construction uses 7-branes in AdS$_{10}$, each of which individually is codimension-2. In some sense, gravity in the GS construction is
localized in only one of the two extra dimensions whereas the second of
the two   extra  directions is compact, although
it is finite-sized  only because of the AdS space. The precise form of the corrections to Newton's 
Law depend on the resolution of the singular geometry at infinity \cite{poppitzponton}.

 Explicitly, one can write the GS metric as

\be
  ds^2 &=& \sigma(\rho) \eta_{\mu\nu} dx^\mu dx^\nu - d\rho^2 - \gamma(\rho) d\theta^2 
\en

\noindent where the 3-brane is located between $\rho=0$ and $\rho<l$.  Inside the 3-brane, the solution is unknown, but outside the 3-brane the solution is

\be
ds^2 &=& e^{-2k\rho}\eta_{\mu\nu} dx^\mu dx^\nu -d\rho^2 -R_0^2 e^{-2k\rho} d\theta^2
\en
  
  Since the warping does not depend
on the sixth GS dimension, for the purpose of finding a consistent metric,
it is best to think of it  as an
additional flat dimension, rather than a warping direction.   This dimension is distinguished from the four infinite
dimensions only by the periodic boundary condition that is imposed. The radius is not a parameter however since
the radius at 
the brane boundary is determined by Einstein's Equations.

  GS put in $T^\mu_{\s \nu} = diag(f_\nu)$ and found two relations,

\be
(\sigma \sigma' \sqrt{\gamma})' &=& -\half \sqrt{-g} \left(G^\rho_{\s \rho} + G^\theta_{\s \theta} \right) \\
(\sigma^2 (\sqrt{\gamma})' )' &=& -\sqrt{-g} \left( G^0_{\s 0} + \frac{1}{4} G^\rho_{\s \rho} -\frac{3}{4} \
G^\theta_{\s \theta} \right)
\label{gsrelations}
\en

\noindent  This in turn led to tension relations for their string-like
solution regardless of details of the brane, since the LHS of both relations above is a total derivative that can be integrated over the brane.  The fact that such a trick was possible depended on the terms $\sigma'$, $\gamma'$ arising in $G^\mu_{\s \nu}$ only in certain linear combinations.  This  apparently  remarkable coincidence becomes even more remarkable as the number of extra dimensions increases and similar relations continue to hold.  We shall find that such relations are not coincidental but rather are guaranteed to exist by the symmetries of the setup.

\section{Holographic Interpretation of Higher Codimension Geometries}

 We will not use the precise form of the holographic description but it
 is helpful to have a qualitative picture of the holographic interpretation
 of the AdS theory   when the codimension is greater than one in order to understand
how these constructions are possible in principle.

 We will first think about the codimension-2 Gherghetta-Shaposhnikov
 example. That case is rather easy to interpret because one of the
 dimensions has periodic boundary conditions, making it act essentially
 like a compact dimension from the perspective of the holographic
 interpretation. At any given value of $z$, the radial coordinate, there is
 only a finite sized circle. Although the circle grows to infinite size in
 coordinate units, it is always finite size due to the AdS warp factor. This
 corresponds to what Ponton and Poppitz found \cite{poppitzponton} when they resolved
 the singularity. The corrections to Newton's Law were never of the form you
 would find with two infinite directions, but corresponded  instead to the corrections you would find with either one
 infinite dimension or no infinite dimension at all (when they cut off the
 singularity).  In both cases, they considered the dual interpretation, which was either a lower-dimensional CFT or a CFT with a cut-off.

 Although it looks quite different, the case of intersecting branes should
 behave holographically as well. The point is that one can again choose a single warping
 direction (in this case, the sum of the directions perpendicular to each
 brane). Again, at each point along this infinite direction, the cross
 section is finite in size. This is because one only sees a sector of
 AdS  due to the boundary intersecting branes.  Clearly, our example, which combines together these ideas, behaves in
 the same manner. There is an effectively compact space fibered
over a ``holographic'' direction.  The dual theory should be a \emph{broken} conformal field theory, since the transverse space is not a fixed size.  It would be interesting to investigate the dual theory to intersecting codimension-1 and codimension-2 branes further, since the theories are distinct from AdS spaces that have already been studied.

\section{Metric}

We choose conventions such that  $\eta_{\mu\nu}$ has signature $(+,-,-,-)$ and

\be
R_{AB} - \half g_{AB} R &=& g_{AB} \Lambda + \frac{1}{M_{10}^8} T_{AB} 
\en

\noindent
 For simplicity, we impose a symmetry in the exchange of any two branes, and assume that the setup is 
azimuthally symmetric  around each individual brane. 
We thus make the ansatz\footnote{More generally, we might want
 to have $dz_i dz_j$ terms, but this will not affect the following discussion.}

\be
ds^2 &=& \sigma(\vec{z})g_{\mu\nu}^{(4)}dx^\mu dx^\nu - \sum_i \left( \xi_i(\vec{z}) dz_i^2 + \beta_i(\vec{z}) dy_i^2 \right)
\en

\noindent where $z_i$ is the direction normal to the $i-$th brane and $y_i$ is its angular direction. 

At this point, we choose to smooth out the string over some arbitrarily small distance $\epsilon$.  The metric inside the brane is unknown and  depends on the distribution of energy-momentum on the thickened brane.   We will be interested in the limit as the thickness of the brane becomes arbitrarily small, but never zero.  A brane with truly vanishing thickness must have tension proportional to its induced metric, so in particular $T^{y_i}_{\s y_i}$ for the $i$-th brane would vanish.  However, a large $T^{y_i}_{\s y_i}$ component is necessary for the stabilization of the extra dimensions. Thus, we are led to take the thickness of the brane arbitrarily small at the end of our calculations, and not before.

 The boundary 
conditions that avoid a singularity at the center of each thickened brane are \cite{papantonopoulos,gregory,navarro} 
\be
\d_{z_1} \sigma &=& 0 \qquad \sigma(0,z_2,z_3) = A(z_2,z_3) 
\label{bdfirst} \\ \nn\\
\d_{z_1} \xi_1 &=& 0 \qquad \xi_1 (0,z_2,z_3) = B(z_2, z_3)\\ \nn\\
\d_{z_1} \beta_1 &=& \sqrt{B(z_2,z_3)} \qquad
\beta_1(\epsilon,z_2,z_3) \sim \epsilon \sqrt{B(z_2,z_3)}
\label{bdlast}
\en

\noindent 
and symmetrically for branes 2 and 3.  The functions $A$ and $B$ have finite first derivatives and are symmetric 
in $z_2 \leftrightarrow z_3$, but otherwise generic.  
 To construct solutions in the bulk, we recall the usual procedure of cutting and pasting $AdS$ space along perpendicular 
codimension-one branes, as in e.g. \cite{inflargedims,origami}.

\be
ds^2 &=& { 1 \over ( k \sum_{i=1}^3 |z_i|+1) ^2 } \left( \eta_{\mu\nu} dx^\mu dx^\nu - \sum_{i=1}^{3} dz_i^2 \right)
\label{nimametric}
\en

\noindent where $z_i$ is the direction perpendicular to the brane $i$, respectively, and $k$ is related to the
 cosmological constant (c.c.) $\Lambda$ in the bulk.  This corresponds to a c.c. 
in the bulk and a tensionful brane at $z_i=0$ for each $z_i$.  We would like something similar to this,
 but with codimension-2 branes.  The metric in the bulk from a single codimension-2 brane, 
from \cite{gs}, is

\be
ds^2 &=& e^{-2k\rho}\eta_{\mu\nu} dx^\mu dx^\nu -d\rho^2 -R_0^2 e^{-2k\rho} d\theta^2
\en

\noindent
where $\theta \in \left[ 0, 2\pi \right]$.  With the change of variables $kz+1 = e^{k\rho}$ and
 $\theta = y/R_0(kz+1)$, this takes the form

\be
ds^2 &=& {1 \over (kz+1)^2} \left( \eta_{\mu\nu} dx^\mu dx^\nu -dz^2 -dy^2 \right)
\label{shaposhmetric}
\en

\noindent
where $y \in \left[ 0, 2\pi R_0\right]$.  For codimension-2 branes, then,
 instead of cutting and pasting AdS along the branes as in (\ref{nimametric}), we 
``wrap'' AdS around the branes as in (\ref{shaposhmetric}):

\be
ds^2 &=& {1 \over (k\sum_i |z_i|+1)^2} \left[ g^{(4)}_{\mu\nu} dx^\mu dx^\nu -\sum_i \left( dz_i^2 +dy_i^2 \right) \right]
\label{metric}
\en

This is conformal to a flat metric ($g_{\mu\nu} = \Omega^2 \eta_{\mu\nu}$) with conformal factor $\Omega \equiv 1/(k\sum_i|z_i|+1)^2$.  From the bulk Einstein equations, we find the parameter 
$k$ is determined from $\Lambda$, the c.c. in the bulk, according to 
$k^2 = -\frac{2}{n(D-1)(D-2)}\Lambda = -\frac{1}{108}\Lambda$, where in this case $D=10$ and $n=3$.

 The Planck scale $M_p$ on the intersection is

\be
M_{p}^2
 &=& M_{10}^8\int_{volume} d^{6}x \Omega^{-2} \sqrt{\Omega^{20}} \nn\\
  &=& M_{10}^8\int \frac{dz_1 dz_2 dz_3 dy_1 dy_2 dy_3}{\left(k \left( |z_1|+ |z_2|+|z_3|\right) + 1 \right)^8} \nn\\
  &=& M_{10}^8\frac{(2\pi R_0)^3}{210 k^3}
\en

We note here that we assume there are three perpendicular
codimension-2 branes. 
 In order for this to be a stable configuration, there must be some stabilization mechanism for the angle between the branes. 
 This angle affects the four-dimensional Planck mass on the intersection and thus acts as a 
Brans-Dicke field.  In order to stabilize any given angle, we
need interactions between the branes. These will in general also contribute to the tensions and affect the tension relations we will find.  This
is also an issue for the codimension-1 branes.
 These are essential issues but we leave them for  later and assume stationary, perpendicular, and non-interacting branes.

 We now derive a very useful formula based on the symmetries of our setup.  Let us consider the general case of a metric that does not depend on some number of directions $X_a$.   For each direction $X_a$ that the metric does not depend upon, we have a killing vector $K^\mu = (\d X_a)^\mu$.
  Its norm is $\sqrt{|K^2|} = \sqrt{|g_{aa}|}$.   Further, if the metric is independent of the direction $X^a$, then the symmetry $X^a \leftrightarrow -X^a$ implies $g_{\mu a} =0$ for $\mu \ne a$.   Thus,

\be
\nabla^2 \log \sqrt{g_{aa}} &=&
  \half \nabla_A \nabla^A \log (K^B K_B ) \nn\\
    &=& \nabla^A \left( { K_B \nabla_A K^B \over K^2 } \right) \nn\\
    &=& - { \nabla^A K^2 \over (K^2)^2 } (K_B \nabla_A K^B) + \frac{1}{K^2}\nabla^A (K_B \nabla_A K^B) \nn\\
    &=& -\frac{2(K_B \nabla^A K^B) (K_C \nabla_A K^C)}{(K^2)^2}
   + \frac{(\nabla^A K_B)(\nabla_A K^B) + K_B \nabla^A \nabla_A K^B}{K^2}
\nn\\
  &=& -\frac{K^A K^B R_{AB}}{K^2} + \frac{1}{(K^2)^2} \left[ (K^2)(\nabla^A K_B)^2 -2((K_B \nabla^B) K_A)^2 \right]
\en

\noindent where in the last step we have used $\nabla_{(A}K_{B)} =0$ and $\nabla_A \nabla_B K^A = R_{AB}K^A$ \s. 
 Since $K^\mu =(\d X_a)^\mu = \delta_a^\mu$, the first term is $-R^a_{\s a}$ and the bracketed term vanishes.  
Thus, for each direction $X_a$ that the metric does not depend upon, we have (no implied sum over $a$)

\be
R^a_{\s a} &=& -\nabla^2 \log \sqrt{g_{aa}}
\label{diveqn}
\en

\noindent 
 Equation (\ref{diveqn}) holds at all points that $K^2 \ne 0$.  In the Newtonian limit in a 4D Minkowski background, the case $x^a=x^0=t$ reduces immediately to $4\pi G \rho =   
\nabla^2 \Phi$, since $\log \sqrt{g_{00}} \approx \half h_{00}=-\Phi$. Eq (\ref{diveqn}) 
is essentially Poisson's equation with the tension $T^\mu_{\s \nu}$ acting as a \emph{linear} 
source for $\log g_{aa}$.  We have not linearized gravity yet; the $g_{AB}$ appearing in equation (\ref{diveqn}) 
is the full background metric.  Thus, by Gauss' law we can extract information about the tension on the brane 
by knowing about the physics away from the brane.  We will use this to our advantage in the following analysis.

 We now understand why in the one-brane case it was possible to find tension relations without knowing the detailed distribution of the energy-momentum tensor on the thickened brane.  Equation (\ref{diveqn}) depends only on the symmetry of the setup, and is true in general whenever the metric does not depend upon a direction $X^a$.  To show more explicitly how this leads to the relations in \cite{gs}, we can rewrite equation (\ref{diveqn}) as

\be
R^a_{\s a} &=& -\frac{1}{\sqrt{-g}} \d_A \left( \sqrt{-g} g^{AB} \d_B \log \sqrt{g_{aa}} \right)
\en

and the relations (\ref{gsrelations}) as

\be
-\d_\rho \left( \sqrt{-g} g^{\rho \rho} \d_\rho \log g_{tt} \right) &=&
  2 \sqrt{-g}  R^{t}_{\s t} \\
- \d_\rho \left( \sqrt{-g} g^{\rho \rho} \d_\rho \log \sqrt{g_{\theta \theta}} \right) &=& \sqrt{-g} R^\theta_{\s \theta}
\en

\section{Finding Tension Relations with Stokes' Theorem}

\subsection{Tension Components}

To find the solution to Einstein's equations,
we need to find the relationship between the bulk energy momentum tensor and the tensor components on the brane.
In the case of codimension-2 branes, this might seem an impossible task since the metric for a string-like defect
changes over the string meaning that in general one deals with a thick defect. If you take the infinitely thin string, 
the metric is discontinuous and physical properties can depend on how the limit is taken 
\cite{traschen}. However, we will see that the tension relationships involve only integrated tension 
as well as a few boundary parameters. This follows from Stoke's theorem applied
to our system.

Let us suppose that the branes have some energy-momentum tensor $T^\mu_{\s \nu} = diag( f_i(\vec{z}))$, 
with $f_0 = \dots = f_3$.

The quantities of interest to us are the tension components, defined as

\be
\mu_a \equiv \int d^6x \sqrt{-g} f_a 
\en

\noindent integrated over all three branes. We can write this suggestively as
\be
\frac{1}{M_{10}^8} \left( \mu_a - \frac{1}{8} \sum_{A=1}^{10} \mu_A \right) &=& \int d^6x R^a_{\s a} 
\en
with no summation over $a$.  Thus, knowledge of a component of $R^\mu_{\s \nu}$, or even just its integral, gives us a relation among the tension components. Equation (\ref{diveqn})  then gives us four tension relations, one each for $a=t,\theta_1,\theta_2,\theta_3$.  In each of those cases, $R^a_{\s a}$ is a total derivative, and thus its integral only depends on the metric at the outside boundary of the brane.  Thus, we should be able to derive four tension relations.  We will see that this is  the case, though the tension relations include a constant that depends on the metric at the center of the branes.  A similar constant appears in the tension relations in the case of a single codimension-2 brane in six dimensions \cite{gs}.  However, in that case, the constant was simply the value of the $g_{00}$ component of the metric at the center of the brane, whereas our constant will be an integral along the centers of the branes.

\subsection{The Centers of the Branes}

 We now encounter a subtlety in applying equation (\ref{diveqn}) to our setup. The essential point is that equation (\ref{diveqn}) only holds when $g_{aa} \ne 0$,  so  at the center of each brane it fails to be true. 
At such points, our   derivation fails because we divide by $g_{aa}=K^2$ in several places.  We would like to know what to replace it with.  The RHS, $-\half \nabla^2 \log K^2$, is easily seen using Gauss' Law to be proportional to $\sum \delta(z_i) \nabla_{z_i} K^2$.  The Ricci tensor is more complicated.  
Consider first a simple example for a thickened string:
\be
ds^2 &=& dt^2 - dz^2 - dr^2 - \beta^2(r) d\phi^2 
\label{straightstring} \\
\beta(r) &=& (l/\gamma)\sin (r\gamma/l) \qquad r<l
\en
where this is matched onto a flat geometry at $r>l$.  A quick calculation gives $R^\phi_{\s \phi} = \frac{\beta''}{\beta}=-\frac{\gamma^2}{l^2}$ and $\int_0^l dr \int_0^{2 \pi} \sqrt{-g} R^\phi_{\s \phi} = 2\pi ( \beta'(l) - \beta'(0) ) = 2\pi(\cos(\gamma) - 1)$.  We could take equation (\ref{diveqn}) literally and convert 

\be
\int_{r<\epsilon} dr d\phi \sqrt{-g} R^\phi_{\s \phi} &=& 
-\int_{r<\epsilon} dr d\phi \sqrt{-g} \nabla^2 \log \beta
\label{badintegral}
\en

\noindent into a surface integral at $r=\epsilon$, giving $2 \pi \cos(\gamma)$.  This clearly conflicts with the correct answer.

  Of course, the reason for the conflicting answer is that we have integrated equation (\ref{diveqn}) over a region including the point $r=0$.        At this point, $g_{\phi\phi}=0$.  The correct way to apply (\ref{diveqn}) to the LHS of (\ref{badintegral}) is to evaluate the contributions from the point $r=0$ and from the points $0<r<\ep$ separately.  We can correctly use (\ref{diveqn}) to turn $\int_{0<r<\ep} dr d\phi \sqrt{-g} R^\phi_{\s \phi}$ into a surface integral.  This surface integral is now over the two boundaries ($r=0$ and $r=\ep$) of the region $0<r<\ep$.  The contribution from the interior boundary is $-2\pi$, exactly the term missing earlier.    We still should include the contribution from the point $r=0$, but this is trivial; its contribution is zero.  The reason is that the metric (\ref{straightstring}) avoids a $\delta$-function singularity at the center of the thickened string as long as $\beta$ satisfies the boundary condition $\beta'(0)=1$\footnote{If $\beta'(0)=1$, the metric is locally Minkowski space near $r=0$.}.  

  The above example contains the essential idea behind the procedure we will apply to our thickened 7-branes.
We chose our boundary conditions (\ref{bdfirst})-(\ref{bdlast}) specifically to avoid a singularity at the center.  Of course, generic boundary conditions at the center of the branes will give rise to singularities.  In section \ref{riemannsingularities}, we derive the form of such singularities in order to demonstrate that our boundary conditions do indeed set them to zero.  Physically, these boundary conditions correspond to the fact that we are dealing with thickened branes, so that the tension is smeared out over a small but finite length. Thus, equation (\ref{diveqn}) holds everywhere except for $r=0$, where the LHS is finite but the RHS is singular.

\subsection{Tension Relations}

In light of the previous discussion, the proper procedure should now be clear.  We are interested in the integral of $R^a_{\s a}$ over the entire brane.  We split this integral up into two regions, $M_{center}$ and $M$, where $M_{center}$ is an arbitrarily small open set around $r=0$. $M$ covers the rest of the brane.  We have argued that the integral over the $M_{center}$ vanishes since our boundary conditions set $R$ to be regular.   $M$ now contains all points on any of the branes except for their centers.   The boundary $\d M$ of $M$ therefore contains both an outer surface $\d M_{outer}$ and an inner surface $\d M_{inner}$.  We will use Stokes' theorem to convert the volume integral over $M$ into a surface integral over $\d M$.  We will find that the surface integral over $\d M_{inner}$ does not vanish in all cases\footnote{This is a distinctly different integral from the volume integral over $M_{center}$.  The integral over $M_{center}$ is a volume integral and only has contributions from the tension inside $M_{center}$.  The integral over $\d M_{inner}$, however, is an integral of the flux through a surface and gets contributions from the tension on the entire brane.}.   
  Take  $\gamma$ to be the induced metric on $\d M$,
and $n^A$ the unit outward normal vector to $\d M$.   Integrating both sides of Einstein's equations
over $M$ gives

\be
\frac{1}{M_{10}^8} (\mu_a - \frac{1}{8} \sum_{A=1}^{10} \mu_A)  &=&
   -\int_M d^6x  \sqrt{-g} \nabla^A \nabla_A \log \sqrt{g_{aa}} \nn\\
   &=&  -\int_{\d M} d^5x \sqrt{-\gamma} n^A \nabla_A \log \sqrt{g_{aa}}
\nn\\
 &=& -3\frac{(2\pi R_0)^3}{56k}-3(2\pi R_0)^3\int dz_2 dz_3\left[
{\sqrt{-\gamma} \over \sqrt{B(z_2,z_3)} } \frac{\d_{z_1}\sqrt{g_{aa}}}{\sqrt{g_{aa}}}\right]_{z_1=0} 
\label{surfaceintegral}
\en

\noindent  For convenience, we define $\mathcal{D}_a\equiv \int dz_2 dz_3\left[
{\sqrt{-\gamma} \over \sqrt{B(z_2,z_3)} }  \frac{\d_{z_1}\sqrt{g_{aa}}}{\sqrt{g_{aa}}}\right]_{z_1=0}$, the surface integral over $\d M_{inner}$ inside the first brane. 
 The boundary conditions (\ref{bdfirst})-(\ref{bdlast}) imply that $\mathcal{D}_0 =0$.  This leaves us with only one unknown integral, $D_{\theta_1} = D_{\theta_2} = D_{\theta_3} \equiv D_{\theta}$.  

After some simplification, (\ref{surfaceintegral}) can be rewritten as follows:

\be
\mu_{\theta_1} &=& \mu_{\theta_2} = \mu_{\theta_3} \equiv \mu_{\theta} \\
\mu_0 &=& \mu_\theta+3(2\pi R_0)^3\mathcal{D}_\theta 
\label{tuningeqn} \\
\frac{1}{M_{10}^8}(\half\mu_0 -\frac{3}{8}\mu_\theta- \frac{1}{8} \sum_{i=1}^3 \mu_{z_i} )  &=& -3\frac{(2\pi R_0)^3}
{56k}
\label{rzero}
\en

 The compactification scale $R_0$ is
 determined from the components of the tension according to eq (\ref{rzero}).  Eq (\ref{tuningeqn}) 
indicates a tuning-condition of the tension components.  Notice that this depends only on the metric 
at the center and exterior surface of the brane, but not on the metric in between. This is similar to
 the well-known behavior of the potential outside a distribution of electric charge.  In GR, though, 
there can be different tensions on a codimension-2 brane which correspond to the same solution
 outside the brane (see \cite{traschen} for a thorough discussion).   The above relations among those 
tensions, however, do not suffer from the same ambiguous behavior.

The metric on the intersection of the branes does not have to be Minkowski space, and we can ask how the above relations change if the intersection is $dS_4$ or $AdS_4$ space with a four-dimensional $\Lambda_{phys}$.   $\Lambda_{phys}$ is defined by  $R^{(4)}_{\s\s\mu\nu} -\half g^{(4)}_{\s\s\mu\nu} R^{(4)} = g^{(4)}_{\s\s\mu\nu} \Lambda_{phys}$.  $\Lambda_{phys}$ can be tuned to zero by tuning the tensions and $R_0$ to satisfy equation (\ref{rzero}).  Instead of tuning $\Lambda_{phys}$ to be
 zero, we can allow it to be small but non-zero, in which case   
the metric will have a more complicated dependence on $\vec{z}$.  In appendix \ref{lambdaphys}, 
we produce the appropriate metric and calculate the modification to (\ref{surfaceintegral}),
 but in fact it can be deduced by dimensional arguments.  As $\Lambda_{phys}$ approaches zero, 
we must recover (\ref{surfaceintegral}) above.  Furthermore, $\Lambda_{phys}$ has units of 
$(mass)^{2}$, and the only other dimensionful quantities around are $k$ and $R_0$.  
By azimuthal symmetry, the new contribution must have the same factor of $(2 \pi R_0)^3$. The only modification in the tension relations is in equation (\ref{rzero}):
\be
\frac{1}{M_{10}^8}(\half\mu_0 -\frac{3}{8}\mu_\theta- \frac{1}{8} \sum_{i=1}^3 \mu_{z_i} )  &=& -3\frac{(2\pi R_0)^3}
{56k}(1-c_\Lambda \frac{\Lambda_{phys}}{k^2} )
\label{clambda}
\en
\noindent where $c_\Lambda$ is some constant $ \sim \mathcal{O}(1)$.

\subsection{Tensions at the Intersection of Branes}

We can now ask, in the limit of arbitrarily thin branes, what equation (\ref{diveqn}) tells us about the tensions where two or more branes intersect.  We make the replacement

\be
f_A(\vec{z}) &\rightarrow& \sum_{i=1}^3 \mu^{(i)} \delta_\ep (z_i) + \sum_{i\ne j} \mu^{(ij)}_A \delta_\ep (z_i)\delta_\ep (z_j) +
\mu^{(123)}_A\delta_\ep (z_1)\delta_\ep (z_2)\delta_\ep (z_3)
\en

\noindent where we are leaving open for the moment the possibility that there is some tension associated with 
the intersections of the branes.  The $\delta_\ep$ functions are, of course, not true $\delta$-functions, but are smeared out over the brane thickness $\ep$, which we take to be arbitrarily small.

  We will see that, for local branes with $\mu_{zz}=0$, the tensions of the intersection will vanish.  One might expect this \emph{a priori}.  In \cite{origami}, 
the author studied two intersecting codim-1 branes and found that the tension on the intersection
 vanished precisely when the branes met at right angles.  To see this explicitly for our case,
 we again use Gauss' law but this time with two (three) of the radial directions $z_i$ at an 
arbitrarily small distance $\epsilon$ to study the intersection of two (three) branes.  

  To study the intersection between two branes, begin by taking a small six-dimensional volume around branes 
1 and 2 defined by

\be
V&\equiv&\left\{ x^\mu \in M \arrowvert \s |z_1| + |z_2| < \ep\right\} \\
\Sigma &\equiv& \d V = \left\{ x^\mu \in M \arrowvert \s |z_1| + |z_2| = \ep\right\}
\en

Define coordinates $\{x_\mu,w,z_3,y_i\}$ on $\Sigma$ with $w=z_2-z_1$.  Then, on $\Sigma$, 
$z_1 = \ep - z_2 = \frac{\ep - w}{2}$, $z_2 = \frac{w+\ep}{2}$.
Since $\gamma_{ij} = \dd{x^\mu}{y^i} \dd{x^\nu}{y^j} g_{\mu\nu}$, where $\gamma$ is the induced metric on $\Sigma$, 
we have $\gamma_{AB}=g_{AB}$ component by component and $\gamma_{ww}=\frac{1}{4}(\xi_1+\xi_2)$.
The normal vector to $\Sigma$ is $n^A = \frac{(\d_{z_1})^A + (\d_{z_2})^A}{\sqrt{\xi_1+\xi_2}}$ and thus the integral
 of $R^{y_1}_{\s y_1}$ over $V$ is

\be
\int_V d^6x \sqrt{g} R^{y_1}_{\s y_1}  
  &=& -\int dz_3 \Omega ^6 \frac{1}{2\sqrt{2}} \left( \d_{z_1} \Omega + \d_{z_2} \Omega \right) \left( 2\pi R_0 \right)^3 \int_{-\ep}^\ep dw 
  \nn\\
  &+& \frac{1}{\sqrt{2}} \int dz_3 A^2 B^2\sqrt{\xi_3}\beta_3 \epsilon
   \int_{-\epsilon}^\epsilon  dw \nn\\
  &\stackrel{\ep\rightarrow 0}{\longrightarrow}& 0
\label{nodeltaproducts}
\en

\noindent 
So $R^{y_1}_{\s y_1}$ contains no product of $\delta_\ep$-functions. $R^0_{\s 0}$ vanishes similarly, and the integrals 
over the triple intersection vanish even more quickly since the volume shrinks faster.  We would expect this 
situation to change if we added a stabilizing potential or interactions between the branes.  Perhaps the least 
complicated correction to this is that, when the branes form oblique angles with each other, the metric should 
have explicit factors of $\cos(y_i)$.  These have implicit jumps at $z_i = 0$, thereby introducing $\delta$-functions 
under the integral of eq  (\ref{nodeltaproducts}).  

The vanishing of (\ref{nodeltaproducts}) implies further that the tension on the triple intersection vanishes. 
 We already know that by conservation of energy, $\mu^{(123)}_{z_i}$ is zero.  Thus, the analogues of eq's (\ref{tuningeqn}) and 
(\ref{rzero}) imply $\mu^{(123)}_0=\mu^{(123)}_\theta=0$.  

\section{Curvature at Singularities}
\label{riemannsingularities}

 We will now derive a formula for $\delta$-function singularities at the origin for spacetimes with rotational symmetry.  We want to know the value of $R^a_{\s a}$ where one of the radial coordinates $z$ vanishes.  At such a point, the metric is degenerate and all values of $y$ correspond to the same point in space-time.  At a non-degenerate point, the Riemann tensor depends on the change in a vector as it is parallel transported 	around a loop with four sides, as in Figure \ref{riemann}.  At a degenerate point, however, there are only three sides.  
We can take $\phi = y/R_0$, so that $\phi \in [0,2\pi]$.  Parallel transporting a vector $v^a$ around such a loop gives, with

\be
\delta v^d &=& \delta_3 + \delta_1 + \delta_2 \\
           &=& - \left( dz \d_z v^d \right)_{(dz/2,d\phi)} + \left( dz \d_z v^d \right)_{(dz/2,0)} + \left( d\phi \d_\phi v^d \right)_{(dz,d\phi/2)}  \\
           &=& \left[ \left( dz \Gamma^d_{zb} v^b \right)_{(dz/2,d\phi)} + \left( -dz \Gamma^d_{zb} v^b \right)_{(dz/2,0)} \right]
              + \left( -d\phi \Gamma^d_{\phi b} v^b \right)_{(dz,d\phi/2)}
             \\
       &=&  \left[ dz d\phi \d_\phi \left( \Gamma^d_{zb} v^b \right)_{(dz/2,d\phi/2)} \right] 
-\left(d\phi \Gamma^d_{\phi b} v^b \right)_{(0,d\phi/2)}
 - d\phi dz \d_z \left( \Gamma^d_{\phi b}v^b\right)_{(dz/2,d\phi/2)} \\
    &=&
       dz d\phi \left( \d_\phi \Gamma^d_{zb} -\d_z \Gamma^d_{\phi b} + \Gamma^d_{\phi e} \Gamma^e_{z b} - \Gamma^d_{z e} \Gamma^e_{\phi b} \right) v^b - dz d\phi \delta(z)  \Gamma^d_{\phi b} v^b
\en

Thus, the Riemann tensor is 

\be
R_{z \phi b}^{\s \s \s d} &=& 
   \left( \d_\phi \Gamma^d_{zb} -\d_z \Gamma^d_{\phi b} + \Gamma^d_{\phi e} \Gamma^e_{z b} - \Gamma^d_{z e} \Gamma^e_{\phi b} \right) 
    -\delta(z)  \Gamma^d_{\phi b} +\Delta_b^{\s d}
\en

\noindent The term $\Delta_b^{\s d}$, which we derive below,  is required in order to rotate the basis vectors $(\d_z)^a$ and $(\d_\phi)^a$ back to their original position.

\FIGURE{
\includegraphics[width=0.8\textwidth]{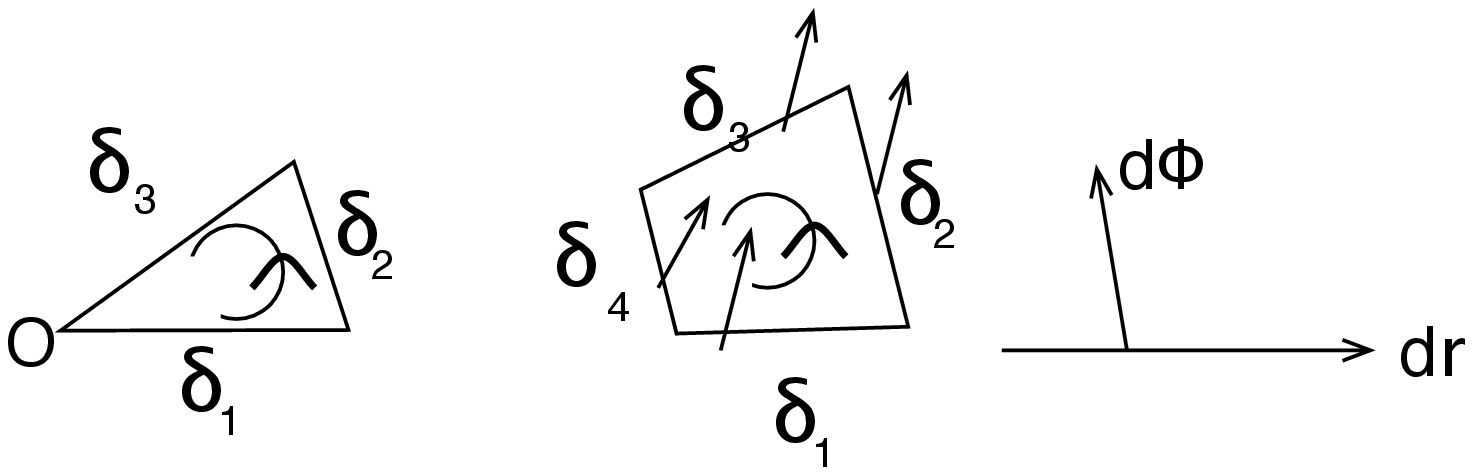}
\caption{A vector $v^a$ is parallel transported around a closed loop.  Usually, the loop will have four sides, two for constant $r$ and two for constant $\phi$, but at the origin there are only three sides.}
\label{riemann}
}

A passive clockwise rotation, to undo the rotation along $\delta_2$, will take

\be
{(\d_z)^a \over \sqrt{|\d_z|^2} } &\rightarrow&
    \cos(d\phi) {(\d_z)^a \over \sqrt{|\d_z|^2} } 
   + \sin(d\phi) {(\d_\phi)^a \over \sqrt{|\d_\phi|^2} } \\
{(\d_\phi)^a \over \sqrt{|\d_\phi|^2} }   &\rightarrow&
    \cos(d\phi) {(\d_\phi)^a \over \sqrt{|\d_\phi|^2} }
   -\sin(d\phi) {(\d_z)^a \over \sqrt{|\d_z|^2} }
\en

Thus, $v^a = \frac{1}{g_{zz}}\left( (\d_z)_b v^b \right) (\d_z)^a
    + \frac{1}{g_{\phi \phi}}\left( (\d_\phi)_b v^b \right) (\d_\phi)^a$ will go to

\be
v^a+ \delta v^a &=& \frac{1}{g_{zz}}\left( (\d_z)_b v^b \right) \left( (\d_z)^a +d\phi (\d_\phi)^a \sqrt{\frac{g_{zz}}{g_{\phi\phi}}} \right) \nn\\
    &+& \frac{1}{g_{\phi \phi}}\left( (\d_\phi)_b v^b \right) 
   \left( (\d_\phi)^a - d\phi (\d_z)^a \sqrt{\frac{g_{\phi\phi}}{g_{zz}}} \right) \nn\\
   &=& v^a - d\phi \frac{1}{\sqrt{g_{\phi\phi}g_{zz}}} \left\{ \left[ (\d_\phi)_b v^b\right] (\d_z)^a - \left[(\d_z)_b v^b\right] (\d_\phi)^a \right\}
\en

\noindent
which in turn implies

\be
\delta v^a &=& dz d\phi \delta(z) \frac{1}{\sqrt{g_{\phi\phi} g_{zz}}}
   v^b \left\{ g_{z b} \delta^a_\phi - g_{\phi b} \delta^a_z \right\} \\
   &\stackrel{def}{=}& 
     dz d\phi v^b \Delta_b^{\s a} 
\en

Thus, the full Riemann tensor is

\be
R_{z \phi b}^{\s \s \s d} &=& 
   \left( \d_\phi \Gamma^d_{zb} -\d_z \Gamma^d_{\phi b} + \Gamma^d_{\phi e} \Gamma^e_{z b} - \Gamma^d_{z e} \Gamma^e_{\phi b} \right)  \nn\\
    &-&\delta(z)  \Gamma^d_{\phi b} 
    -\delta(z)  \frac{1}{\sqrt{g_{\phi\phi} g_{zz}}}
  \left\{ g_{\phi b} \delta^d_z - g_{z b} \delta^d_\phi \right\}
\label{newriemann}
\en

In an appendix, we apply this to \{ the straight string metric
\be
ds^2 &=& dt^2 - dz^2 -dr^2 - \beta^2(r) d\phi^2
\en
and \} we find that the contribution at $r=0$ is

\be
\int d\phi \sqrt{-g} R^{\phi}_{\s \phi} &=& 2\pi \delta(r) (\beta' -1)
\label{straightstringtension}
\en

 Now, we can say more precisely why we cut out the center of the thickened branes.   Equation (\ref{diveqn}) is true except at the center of each brane, where the singular part of the LHS can be calculated from equation (\ref{diveqn}).  The singular pieces of the LHS and of the RHS of (\ref{diveqn}) will not in general be equal.  Thus, in order to take advantage of Stokes' theorem, we divided up the brane into two parts: away from the center, where (\ref{diveqn}) applies, and at the center, where it does not.  We then evaluated the integral over the former piece using Stokes' theorem and over the latter piece using equation (\ref{newriemann}).  We claimed that the contribution from the latter piece should vanish in our case.  We can now see this explicitly.  

\be
R_{z_1 y_1 z_1 }^{\s \s \s \s \s \s y_1} &=& -\delta(z_1) \left[
   \Gamma^{y_1}_{z_1 y_1} +
\frac{1}{\sqrt{g_{y_1 y_1} g_{z_1 z_1}}} \left( -g_{z_1 z_1} \right) \right] 
\en

\noindent so

\be
\sqrt{-g} R^{z_1}_{\s z_1} &=& -\delta(z_1) \beta_2 \beta_3 \sigma^2 \sqrt{\xi_2 \xi_3/\xi_1} \left[ \d_{z_1} \beta_1 - \sqrt{\xi} \right]
\en

\noindent which vanishes under the boundary conditions (\ref{bdfirst})-(\ref{bdlast}).  The other components of $R^\mu_{\s \nu}$ vanish similarly.

We have so far only discussed calculating the singularity of $R^\mu_{\s \nu}$ at the origin for the purpose of setting it to zero, but it is useful more generally.  For instance, consider the infinitely thin, straight string.  It gives rise to a geometry that is locally flat except for a deficit angle.   In this case, there is an actual singularity at $r=0$, and the interpretation of $R^\mu_{\s \nu}$ at $r=0$ is quite different.  Rather than enforcing some boundary conditions, the singularity is the actual distributional energy-momentum tensor of the string.  Explicitly, take the metric (\ref{straightstring}) with
\be
\beta(r)&=& (1-4 G \mu)r
\en
This corresponds to a deficit angle $\Delta = 8 \pi G \mu$ and an energy-momentum tensor $T^\rho_{\s \sigma} = \mu \delta(r) diag(1,1,0,0)$.  For this simple example, we can linearize gravity to calculate $R^\mu_{\s \nu}  = -8 \pi G\mu \delta(r) (0,0,1,1)$ explicitly \cite{vilenkin}. 
In this case, equation (\ref{straightstringtension}) gives $R^\phi_{\s \phi} = -8 \pi G \mu \delta(r)$, as it must.

\section{Localization}

In order to check that our construction localizes gravity to the 4-dimensional intersection of the branes, we need to consider the spectrum of graviton modes.
We can find the effective potential for the graviton $h_{\mu\nu}$ in the usual way,
by plugging $h_{\mu\nu} = \Omega^{-(D-2)/2}(\vec{z}) e^{ip\cdot x} \tilde{h}_{\mu\nu}(\vec{z})$ into

\be
\frac{1}{\sqrt{g}} \d_A \left( \sqrt{g} g^{AB} \d_B h_{\mu\nu} \right) &= &0
\label{grlaplacian}
\en

\noindent to get a Schrodinger wave equation for the graviton:

\be
\left( \d^2 +m^2 -V(\vec{z}) \right) \tilde{h} &=& 0
\en

\noindent
where $V(\vec{z})=12 \frac{(\d_z \Omega)^2}{\Omega^2}+ 4 \frac{\d_z^2 \Omega}{\Omega}$ in the bulk, and we are using
 $\tilde{h}$ to represent any of its components.  Now, in the bulk $\d_{z_i} \Omega$ and $\d_{z_i}^2 \Omega$ are 
trivially $-k \Omega^2$ and $2k^2 \Omega^3$, respectively.  The value of $\d_z^2 \Omega$ at the branes is a little
 more subtle, since

\be
\dd{^2 |z_i|}{z_i^2} &=& 2 \pi R_0 \delta(z_i)
\en

\noindent
To see this, compare the straightforward expansion $\nabla^2 \Omega = -3k^2D\Omega + k\sum_{i=1}^3 \dd{^2 |z_i|}{z_i^2}$ 
with  a  routine Gauss' Law computation of the integral of $\nabla^2 \Omega$:

\be
\int_{|z_1|=\ep} \sqrt{g} \nabla^A \nabla_A \Omega dz_1 dy_1 &=&
\int_{|z_1|=\ep} \sqrt{\gamma} n^A \nabla_A \Omega dy_1 
= -\Omega^{D-1} \frac{1}{\Omega} (-\Omega^2 k) \dd{|z_1|}{z_1} 2\pi R_0  
\stackrel{\ep \rightarrow 0}{\longrightarrow} 2\pi R_0 k \nn
\en

\noindent
and thus

\be
V(\vec{z}) &=& {60k^2 \over (k(\sum_i |z_i|)+1)^2} -{8\pi R_0 k \over k\sum_i |z_i| +1} \sum_i \delta(z_i)
\en

This is a volcano potential along each of the branes.    The form of the volcano potential is itself enough to indicate 4-D gravity on the intersection.  Qualitatively, the potential is nearly flat in the bulk, rises sharply as it 
approaches any of the branes, and turns and falls into an infinitely deep potential well at the brane itself. 
This potential well is enough to support our single bound graviton mode.  All the light modes will be exponentially 
damped as they tunnel through the potential barrier around the branes, leaving gravity essentially four-dimensional 
at low energies. The more energetic the mode, the less tunneling it takes to reach the brane, and at high enough energy 
(roughly, at about $7.7k$, the peak of the potential) an observer would see ten-dimensional gravity recovered.  
The $\delta$-functions are merely enforcing a boundary condition.  To derive this boundary condition, we can isolate 
the $\delta$-function type terms in 
$\d_z^2 \tilde{h} = \dd{^2|z|}{z^2} \d_{|z|}\tilde{h}+ \dd{|z|}{z}^2 \d_{|z|}^2 \tilde{h}(|z|) = 2\pi R_0 \delta^{(2)}(z) 
\d_{|z|} \tilde{h} + \d_{|z|}^2 \tilde{h}(|z|)$. The boundary condition is therefore

\be
-4\frac{k}{k(|z_2|+|z_3|) + 1} \tilde{h}|_{z_1=0} &=& \d_{|z_1|}\tilde{h}|_{z_1=0}
\label{gravbc}
\en

\noindent and symmetrically for $z_2,z_3$.  We note that the zero mode 
$\tilde{h}_0 = \Omega^{4}(\vec{z})$ identically satisfies these boundary conditions.  This matching is trivial 
from the fact that $\tilde{h}_0$ corresponds to $h_{\mu\nu} = const$, which clearly satisfies eq (\ref{grlaplacian}). 

\section{Conclusion}

In this paper, we have constructed 4D gravity in ten dimensions out of 7-branes, essentially 
as the intersection between three copies of RSII.  Due to the symmetry of the setup, we can 
generalize previous methods of relating the brane tension to  the curvature of spacetime outside the branes
(e.g. \cite{gs,navarro}) to extract information about the brane intersections. 
 As usual, there is a volcano potential with an exactly solvable zero mode, as well as a continuum of massive modes.

In the course of our analysis we have derived some interesting features of Einstein's equations. We found that whenever the metric does not depend on a coordinate $x$, the corresponding component of the Ricci tensor $R^x_{\s x}$ is a total derivative $-\half \nabla^2 \log |g_{xx}|$.  We also found a formula for curvature singularities arising from the origin in polar coordinates.

We note that although we will demonstrate that gravity can be localized on the intersection of 7-branes, the filling fraction of the intersection will not in general be the most likely place for our universe to form if the branes forming it are infinite in extent.  However, it could be competitive if they loop around and form loops or some similar configuration, since such a setup would act like a 3-brane on larger scales.  This requires further study which we leave to further work.  Here we show only that the scenario of Ref. \cite{relaxing} can consistently include four-dimensional gravity, even when no dimensions are compactified.

\s

\s

\acknowledgments

\s

We would like to thank  Nemanja Kaloper, Andreas Karch, Juan Maldacena, Veronica Sanz-Gonzalez, Yael Shadmi, Jesse Thaler, and Toby Wiseman for useful discussions.    The work here was supported in part by NSF Award PHY-0201124.

\appendix

\section{$\delta(r)$ Contributions to Curvature}
\label{deltacurv}

We can calculate the singularity from equation (\ref{newriemann}) for straight string metrics: 

\be
ds^2 &=&  dt^2 - dz^2 - dr^2 -\beta^2(r) d\phi^2
\en

\noindent
The non-vanishing Christoffel symbols are

\be
\Gamma^r_{\phi\phi} &=& -\beta \beta' \\
\Gamma^\phi_{\phi r} &=& \beta' / \beta
\en

\noindent
The curvature is thus

\be
R_{r r} &=& R_{r \phi r}^{\s \s \s \phi} \\
       &=& -\delta(r) \left( \Gamma^\phi_{\phi r} + \frac{1}{\beta} \left( -g_{rr} \right) \right) \\
   &=& -\delta(r) \left( \frac{\beta'}{\beta} - \frac{1}{\beta} \right) \\
R_{\phi \phi} &=& -R_{r\phi \phi}^{\s \s \s r}  \\   
   &=& \delta(r) \left( \Gamma^r_{\phi \phi} + \frac{1}{\beta} \left( \beta^2 \right) \right) \\
  &=& \delta(r) \left( -\beta \beta' + \beta \right) 
\en

\noindent
and thus

\be
\int d\phi \sqrt{-g} R^r_{\s r} &=& 2\pi \delta(r) \left( \beta' - 1 \right) \\
\int d\phi \sqrt{-g} R^\phi_{\s \phi} &=&  2\pi \delta(r) \left( \beta' - 1 \right) 
\en

This is what one finds integrating $R$ explicitly over a thickened string with $\beta'=1$ at the center of the string and $\beta'$ above being the value at the edge of the thickened string.  Notice that the above terms vanish for minkowski space, $\beta(r) = r$, as they must.

\section{$\Lambda_{phys} \ne 0$ Contribution to Tension Relations}
\label{lambdaphys}

The metric for $n$ intersecting codimension-two branes in AdS$_{4+2n}$ with AdS$_4$ on the intersection is

\be
ds^2 &=& {L^2 \over (c(\sum_i z_i) + L)^2} 
    \left(\Delta(\vec{z}) g_{\mu\nu} dx^\mu dx^\nu - \sum_i dz_i^2 
          + \sum_{ij} { |\Lambda_{phys}| z_i z_j dz_i dz_j \over \Delta(\vec{z})} \right.
\nn\\
   &-& \left. \sum_i \left( 1 -  \frac{L|\Lambda_{phys}|}{c}\frac{\sum_j z_j}{n}
   \pm a_n \frac{L|\Lambda_{phys}|}{c}{( n z_i - \sum_j z_j) \over n} \right)^2 dy_i^2 \right) \\
  \Delta(\vec{z}) &=& 1 + |\Lambda_{phys}| \sum_j z_j^2 \\
   a_n &=& \sqrt{1+\frac{nc^2}{|\Lambda_{phys}|L^2}}
\en

The c.c. on the intersection is $-|\Lambda_{phys}|$ and the c.c. in the bulk is $\Lambda=-\half(D-1)(D-2)(n\frac{c^2}{L^2}+|\Lambda_{phys}|)$. The warp factor for the 4D metric and for the angular directions have been normalized to unity on the intersection.  In the limit $\Lambda_{phys}\rightarrow 0$, we recover the Minkowski solution (\ref{metric}). To calculate the modified tension relations, we once again use (\ref{diveqn}), as follows.  The induced metric $\gamma$ on the hyper-cylinder surrounding the first brane is diagonal except for the block with the normal directions $z_i$.  This block has eigenvalues $\{1,1, \dots, 1, {1+z_1^2 |\Lambda_{phys}| \over 1+ |\Lambda_{phys}| \sum_i z_i^2 } \}$, so we can easily evaluate $\det (-\gamma)$.  Taking $n=3$,

\be
\int d^{6}x \sqrt{-g} R^0_{\s 0} &=& 
    - (2\pi R_0)^3 \int_0^\infty dz_2 dz_3 \sqrt{-\gamma}  n^{z_1} \d_{z_1} \log \sqrt{g_{00}} \nn\\
   &=& - (2\pi R_0)^3 \frac{c}{L}\int_0^\infty dz_2 dz_3 \left( { L \over c (z_2 + z_3) +L} \right)^9  \nn\\
   &\times& \left( 1+ |\Lambda_{phys}| (z_2^2 + z_3^2) \right)^{3/2} \nn\\
   &\times& \left( 1-\frac{L}{c}|\Lambda_{phys}| {z_2+z_3 \over 3} 
       \pm a_n \frac{L}{c}|\Lambda_{phys}| {-z_2-z_3 \over 3} \right) \nn\\
   &\times& \left( 1-\frac{L}{c}|\Lambda_{phys}| {z_2+z_3 \over 3} 
       \pm a_n \frac{L}{c}|\Lambda_{phys}| {2z_2-z_3 \over 3} \right) \nn\\
   &\times& \left( 1-\frac{L}{c}|\Lambda_{phys}| {z_2+z_3 \over 3} 
       \pm a_n \frac{L}{c}|\Lambda_{phys}| {-z_2+2z_3 \over 3} \right) 
\label{lambdatr}
\en

The large-$z_i$ contribution to the integral is negligible since the integrand drops faster than $\frac{1}{(z2+z3)^2}$.  Thus, in the small-$\Lambda_{phys}$-limit, we can neglect terms of order $\mathcal{O}(\Lambda_{phys}^2)$. Then, (\ref{lambdatr}) is 

\be
\int_{brane \s 1} d^{6}x \sqrt{-g} R^0_{\s 0} &=& -(2\pi R_0)^3 \left( \frac{1}{56k} - \frac{|\Lambda_{phys}|}{420k^3} \right)
\en

\noindent and $c_\Lambda = 2/15$ in eq (\ref{clambda}).

There is another way that one might expect $\Lambda_{phys}$ to enter the tuning relations.  Aside from the solution in the bulk depending implicitly on $\Lambda_{phys}$, the metric on the brane $g^{(4)}_{\mu\nu}$ certainly depends on $\Lambda_{phys}$.  Consequently, $R^{(4) \mu}_{\s\s\s \nu} - \half \delta^\mu_\nu R^{(4)} = \Lambda_{phys} \delta^\mu_\nu$ will contribute to the total $R^A_{\s B} - \half \delta^A_B R$.  However, when we turn to our tension relations, everything is integrated over the branes, whose thickness is only $\epsilon$.  The tension components $f_\mu$ are inversely proportional to $\epsilon$ whereas $\Lambda_{phys}$ is not.  Since $\epsilon$ is very small, this particular contribution from $\Lambda_{phys}$ will be negligible.

\end{document}